# Electrodynamics of magnetoelectric media and magnetoelectric fields

E. O. Kamenetskii

Microwave Magnetic Laboratory,
Department of Electrical and Computer Engineering,
Ben Gurion University of the Negev, Beer Sheva, Israel

**Abstract**
The relationship between magnetoelectricity and electromagnetism is a subject of a strong interest and numerous discussions in microwave and optical wave physics and material sciences. The definition of the energy and momentum of the electromagnetic (EM) field in a magnetoelectric (ME) medium is not a trivial problem. The question of whether electromagnetism and magnetoelectricity can coexist without an extension of Maxwell's theory arises when we study the effects of EM energy propagation and consider group velocity of the waves in a ME medium. The energy balance equation reveals unusual topological structure of fields in ME materials. Together with certain constraints on the constitutive parameters of a medium, definite constraints on the local field structure should be imposed. Analyzing the EM phenomena inside a ME material, we should answer the question: what kind of the near fields arising from a sample of such a material can we measure? Visualization of the ME states requires an experimental technique that is based on an effective coupling to the violation of spatial as well as temporal inversion symmetry. To observe the ME energy in a subwavelength region, it is necessary to assume the existence of first-principle near fields – the ME fields. These are non-Maxwellian near fields with specific properties of violation of spatial and temporal inversion symmetry. A particular interest to the ME fields arises in studies of metamaterials with "artificial-atoms" ME elements.

**I. INTRODUCTION**

In classical condensed-matter electrodynamics, in frequency regions of transparency of the EM-wave propagation in matter, one can introduce the notion of an internal-energy density in dynamic fields in the same sense as in electrostatic and magnetostatic structures [1]. Can we introduce the notion of an internal-energy density in alternating fields in ME medium in the same sense as in magnetoelectrostatic structures? In studies of the propagation of EM waves in ME-media, the concept of internal-energy density is a special question. The ME-material statics and ME-material dynamics (electrodynamics) appear as physically two different behaviors. This makes the problems of interaction of ME structures with external EM fields very nontrivial.

In a case of magnetoelectrostatics, the external electric and magnetic fields are mutually uncoupled. ME effect is described by an expansion of the free energy as a power series in the electric and magnetic fields. A term of the static ME energy density $W_{ME}$ is introduced together with terms of the electrostatic and magnetostatic energy densities [1 – 4]. In spiral multiferroic materials, expansion of microscopic current densities, induced by spins, results in appearance of static toroidal moments of magnetization. The toroidal moment is a vortex-like head-to-tail arrangement of magnetic moments. It appears as a result of spontaneous breaking of both spatial inversion and time reversal symmetry and has been identified as the antisymmetric component of the ME tensor. Introduction of an extended toroid characteristic in condensed matter is the model of isolated point toroid dipoles. In a system with a total toroid moment $\vec{T}$, the free-energy functional contains the contribution $W_{ME} \propto \vec{T} \cdot (\vec{E} \times \vec{H})$ [5 – 7]. Evidently, in structures with magnetic vortices, the toroidal moment could be controlled by the cross product of the electric field and the magnetic field strength. It was realized also that the energy term could be written as $W_{ME} \propto \int \vec{r} \times (\vec{E} \times \vec{H}) \cdot M(\vec{r}) d^3\vec{r}$, where $M(\vec{r})$ is the magnetization [8, 9]. While the toroid moment in media can be excited by taking advantage of the cross product of the electric and a magnetic fields, the creation of such a regime of inhomogeneous field structure in a local ME-effect region, is by no means a trivial problem.

Contrary to magnetoelectrostatics, where the external electric and magnetic fields are mutually uncoupled, in a ME dynamic regime, the *E* and *H* fields are linearly coupled by Maxwell's equations. In these equations, spatial derivatives of *E* are proportional to the time derivative of *H* and vice versa. For a given frequency $\omega$, the scale of the field space variation is the EM wavelength $\lambda$ in a medium. Since both in microwaves and optics, $\lambda$ is much greater than the size of inhomogeneity in the microscopic structure of ME material, it is usually assumed that one can solve electrodynamics problems my standard ways considering monochromatic wave propagation in a dispersive medium with local constitutive parameters. In the constitutive relations, the dynamical response of the first-order ME coupling is described by the dynamical ME tensor in addition to the conventional terms of dynamical dielectric and magnetic susceptibilities. The dynamic susceptibilities can be obtained from the magnetic and electric equations of motion. By combining Maxwell's equations with the ME dynamical-response parameters, different ME electrodynamics problems are formulated. There are, for example, studies of optical effects in ME materials [10 – 13], microwave ME phenomena of skyrmions [14], ME optical waveguide modes [15], ME polaritons [16, 17], and canonical quantization of macroscopic electrodynamics in a ME medium [18].

However, the question, whether electromagnetism and magnetoelectricity can coexist without an extension of Maxwell's theory, arises when we study the effects of EM *energy propagation* and consider *group velocity* of the waves in a ME medium. To derive the EM energy density it is insufficient to consider pure monochromatic fields since no accumulation of the electromagnetic energy takes place in this case. For these effects, we need analyzing of a *quasimonochromatic* regime with the field amplitudes slightly perturbed in time and space [1, 19]. At such a regime of slight space-time "vibrations" of the field amplitudes, the energy balance equation reveals unusual *topological structure* of the local fields in ME materials [20]. These local fields, called *ME fields*, are subwavelength-domain fields with specific properties of violation of spatial and temporal inversion symmetry.

Far-field retrieved EM material parameters frequently retain nonphysical values, especially at the frequencies in the vicinity of the material resonances, where most interesting features are expected. These aspects are very important for probing the ME-material dynamic parameters. The near field structure of such a probe should violate the spatial and temporal inversion symmetries. The cross product $\vec{E} \times \vec{H}$ is related to the EM power-flow density and the term $\vec{r} \times (\vec{E} \times \vec{H})$ is the total angular momentum of an EM field [21 – 25]. It means that to extract local parameters of a ME material with the topological-moment properties, we need a vortex-like arrangement of Poynting vectors in *subwavelength* helical-resonance system. Creation of these twisted (chiral) inhomogeneous EM fields in a local (subwavelength) region is a non-trivial problem [26]. ME power-flow vortices hold energy in its rotation. For an EM-wave flow, such ME vortices are a type of topological defect. Since subwavelength power-flow circulations are quantized, the ME vortices should be quantized vortices. It is worth noting also that in a subwavelength region, ME material contains a dynamic axion field $\theta(t, \vec{r})$. This gives an additional term of the Lagrangian density, $L_\theta \propto \theta \vec{E} \cdot \vec{B}$, which couples electric and magnetic fields. Pseudoscalar term $\theta \vec{E} \cdot \vec{B}$ provides ME coupling in topological insulators and antiferromagnet chromia [27 – 30]. Creation of proper field structures conjugate with the toroidal-order material structures is a crucial, intensely debated point in experimental investigations. The coupled electric and magnetic ordering in ferroelectromagnets, accompanied by the formation of domains and domain walls, were observed at low frequencies by using different techniques of voltage control of magnetism [31 – 33]. At the same time, the question of local-field probing of ME material at microwave and optical frequencies is yet unclear. Microwave responses resulting from ferrotoroidic states $\vec{T} = \vec{P} \times \vec{M}$ observed experimentally in [34 – 37], do not clarify the question on a local field (ME field) structure at a dynamic regime.

Uniqueness of the ME fields can be shown by analyzing vacuum near fields originated from a scatterer made of a ME crystalline material. In this connection, it is worth noting that in a case of usual (non-ME) material structures one can distinguish two kinds of the EM near fields: (a) near fields originated from EM wave resonances and (b) near fields originated from dipole-carrying resonances. The former fields, abbreviated as EM NFs, are obtained based on the full-Maxwell-equation solutions with use of Mie theory [38]. The latter fields, abbreviated as DC NFs, are observed when the electric or magnetic dipole-carrying oscillations (such, for example, as surface plasmons [39 – 41] and magnons [42 – 44]) take place. Notably, in accordance with Mie theory one can observe EM NFs with magnetic responses originated from small nonmagnetic dielectric resonators, both in microwaves [45] and optics [46 – 48]. In a case of DC NFs, strong coupling of EM waves with electric or magnetic dipole-carrying excitations, called polaritons, occur [42, 49]. There is the avoided-crossing coupling between the photon and the dipolar oscillation. Semiclassically, polaritons are described using Maxwell equations and constitutive relations that include the frequency dependent response functions. Quantum mechanically, polaritons are described as hybrid collective excitations that are linear superpositions of matter collective excitations and a photons. There are the effects of interaction between real and virtual photons. Importantly, the spatial scale of the DC NFs is much smaller than the spatial scale of the EM NFs in the same frequency range. Due to the strong coupling of EM waves with dipole-carrying excitations and temporal dispersion of the material, polaritons display enhanced field localization to surfaces and edges. The properties of vacuum near-fields originated from a small non-ME (dielectric or magnetic) sample become evident when this sample has sizes significantly smaller than the EM wavelength *in all three spatial dimensions*. The matter of fact is that near such a scatter we can only measure the electric $E$ or the magnetic field $H$ with accuracy. As volumes smaller than the wavelength are probed, measurements of EM energy become uncertain, highlighting the difficulty with performing measurements in this regime. There is Heisenberg's uncertainty principle binding $E$ and $H$ of the EM wave [39, 50].

Talking now about the near fields of ME-material structures – the ME fields – we are dealing with the effects of strong coupling of EM waves with specific dipole-carrying excitations, called electromagnons. The electromagnons are considered as fundamental excitations that exhibit both electric and magnetic dipole moments [14, 16, 17, 51, 52]. In case of a subwavelength ME sample, the near-field structures are dominated by *two types* of the fields: the electric and magnetic fields, which are *mutually coupled* due to the material properties. When the violation of the invariances under space reflection parity and time inversion are necessary conditions for the emergence of the ME effect, the same symmetry properties should be observed for the near fields – the ME near fields. Due to PT symmetry of ME crystals, the ME near fields of a subwavelength sample should be characterized by a certain pseudoscar parameter. Evidently, such a near-field structure – the ME-field structure – is beyond the frames of the Maxwell theory description [21].

In this paper, we use the Poynting theorem to analyze the properties of ME fields – the local fields in ME materials. We show that both the average energy density and dissipative losses are defined not only by the ME-medium constitutive parameters, describing specific dipole-carrying excitations, but also by particular topological structures of the fields. These topological structures are revealed through constraints imposed to the field amplitudes. We argue that ME energy in EM systems can appear due to local (subwavelength) fields rotated adiabatically in unit cells of ferrotoroidic domains.

ME fields can be considered as quantized states of dipole-dipole magnons in subwavelength domains of ME materials. Such quantized states are well modeled by spectral characteristics of magnetic-dipolar-mode (MDM) oscillations [or magnetostatic (MS) oscillations] observed in confined ferromagnetic-insulator structures. Quantized ME fields arising from nonhomogeneous ferromagnetic resonances with spin-orbit effect, suggest a conceptually new microwave functionality for material sensing. Visualization of the ME states requires an experimental technique that is based on an effective coupling to the violation of spatial as well as temporal inversion symmetry. To observe the local properties of ME materials, we propose the use of special subwavelength probes with ME near fields. The sensor is based on a subwavelength ferrite-disk resonator with MDM oscillations [53, 54].

## II. THE POYNTING THEOREM IN A ME MEDIUM AND THE ME FIELDS

The dynamical processes in a medium with a linear ME effect we describe with use of the integral-form constitutive relations (ICR). These are the linear response functions introduced as [10, 20]

$$D_i(t,\vec{r}) = (\varepsilon_{ij} \circ E_j) + (\xi_{ij} \circ H_j), \tag{1}$$

$$B_i(t,\vec{r}) = (\zeta_{ij} \circ E_j) + (\mu_{ij} \circ H_j). \tag{2}$$

The integral operators on the right side of these expressions have the form similar to the integral operator:

$$(\varepsilon_{ij} \circ E_j) = \int_{-\infty}^{t} dt' \int d\vec{r}' \varepsilon_{ij}(t,\vec{r},t',\vec{r}') E_j(t',\vec{r}'). \tag{3}$$

In Eqs. (1), (2), the causality principle (that is, the fields $D$ and $B$ at the time $t$ are defined by the fields $E$ and $H$ at the time $t' \leq t$) is taken into account [1, 19, 55].

The kernels of the operators in the above ICRs are responses of a medium to the $\delta$-function electric and magnetic fields. So convergence of integrals in the ICRs can be proven if one shows a physical mechanism of influence of short-time and short-space *quasistatic* interactions on the medium polarization properties. When a ME medium is time invariant and spatially homogeneous, the ICRs have a temporal and space convolution form. In such a case, we have a temporary and spatially dispersive ME medium with constitutive parameters satisfying the long-wavelength (quasistatic) limit.

$$\vec{D}(\omega,\vec{k}) = \ddot{\varepsilon}(\omega,\vec{k})\vec{E} + \ddot{\xi}(\omega,\vec{k})\vec{H}, \tag{4}$$

$$\vec{B}(\omega,\vec{k}) = \ddot{\zeta}(\omega,\vec{k})\vec{E} + \ddot{\mu}(\omega,\vec{k})\vec{H}. \tag{5}$$

Such a quasistatic limit takes place when

$$\vec{\vec{\varepsilon}}(\omega,\vec{k})\big|_{|\vec{k}|\to 0} \to \vec{\vec{\varepsilon}}(\omega), \qquad \vec{\vec{\xi}}(\omega,\vec{k})\big|_{|\vec{k}|\to 0} \to \vec{\vec{\xi}}(\omega), \qquad (6)$$

$$\vec{\vec{\zeta}}(\omega,\vec{k})\big|_{|\vec{k}|\to 0} \to \vec{\vec{\zeta}}(\omega), \qquad \vec{\vec{\mu}}(\omega,\vec{k})\big|_{|\vec{k}|\to 0} \to \vec{\vec{\mu}}(\omega). \qquad (7)$$

We can take advantage of the power-series expansion of the constitutive tensors over wave vector $\vec{k}$ in the region near $|\vec{k}| = 0$. That is, the effects of spatial dispersion are considered as small-order effects, similar to the approach used for isotropic and anisotropic media [1, 10, 19, 20, 55, 56]. It can be assumed that in such a ME structure one observes an effect of an interaction between the quasistatic electric and quasistatic magnetic fields.

Suppose that in the frequency regions of the transparency of the electromagnetic-wave propagation in a ME matter, we can introduce the concept of an internal-energy density in alternative fields in the same sense as it magnetoelectrostatic structures. The medium is described by constitutive relations (1), (2). To derive the energy balance equation, we will consider propagation of quasimonochromatic EM waves in ME medium. The fields are expressed as $\vec{E} = \vec{E}_m(t,\vec{r})\, e^{i(\omega t - \vec{k}\cdot\vec{r})}$ and $\vec{H} = \vec{H}_m(t,\vec{r})\, e^{i(\omega t - \vec{k}\cdot\vec{r})}$, where complex amplitudes $\vec{E}_m(t,\vec{r})$ and $\vec{H}_m(t,\vec{r})$ are time and space smooth-fluctuation functions: $\left|\left(\omega^{-1}\frac{\partial}{\partial t}\right)E_{m_i}\right| \ll E_m$, $\left|\left(k^{-1}\nabla\right)E_{m_i}\right| \ll E_m$ and $\left|\left(\omega^{-1}\frac{\partial}{\partial t}\right)H_{m_i}\right| \ll H_m$, $\left|\left(k^{-1}\nabla\right)H_{m_i}\right| \ll H_m$.

Considering corresponding product relations for the Fourier components, we represent every term in the energy balance equation EM waves in a ME medium as a contraction of two tensors: a constitutive tensor (or its derivative) and a dyadic product of the electromagnetic-field vectors. Taking into account that any tensor of the second rank can be written as a sum of Hermitian and anti-Hermitian tensors, it was shown [20] that an energy-conservation law (Poynting's theorem) for time averaged quantities in a ME medium can written as

$$-\nabla\cdot\langle\vec{S}\rangle = \langle Q(\omega,t)\rangle + \langle P\rangle. \qquad (8)$$

Here $\langle\vec{S}\rangle$ is a Poynting vector, $\langle Q(\omega,t)\rangle$ is a term related to the energy accumulation, and $\langle P\rangle$ describes dissipative losses. It was shown [20] that Eq. (8) can be reduced to a form of a continuity equation (the standard-form energy balance equation for the time-averaged values of periodic functions with the period $2\pi/\omega$)

$$-\nabla\cdot\langle\vec{S}\rangle = \frac{\partial\langle W\rangle}{\partial t} + \langle P\rangle \qquad (9)$$

only when the following constraints are imposed to slowly time-varying amplitudes:

$$E_{m_i}^*(t)\frac{\partial H_{m_j}(t)}{\partial t} = H_{m_j}(t)\frac{\partial E_{m_i}^*(t)}{\partial t}. \qquad (10)$$

It means that one can introduce a notion of an average stored energy $\langle W \rangle$ and consider the energy transport in a ME medium only if complex envelopes satisfy Eq. (10). In such a case, the average energy density is expressed as

$$\langle W \rangle = \frac{1}{4}\left\{\frac{\partial(\omega \varepsilon_{ij}^h)}{\partial \omega}E_i^* E_j + \frac{\partial(\omega \mu_{ij}^h)}{\partial \omega}H_i^* H_j + \frac{\partial\left[\omega(\zeta_{ij}^h + \xi_{ij}^h)\right]}{\partial \omega}\left(H_i^* E_j\right)^h + \frac{\partial\left[\omega(\zeta_{ij}^{ah} - \xi_{ij}^{ah})\right]}{\partial \omega}\left(H_i^* E_j\right)^{ah}\right\}, \quad (11)$$

where superscripts *h* and *ah* mean, respectively, "Hermitian" and "anti-Hermitian". In equation (11), the first two terms in the right-hand side describe the energy density of an "ordinary" dielectric/magnetic temporary-dispersive medium, whereas the last two terms characterize the ME energy density, $\langle W_{ME} \rangle$. For quasimonochromatic fields in a weakly absorbing medium the dissipative losses are described by the relation [20]:

$$\langle P \rangle = \frac{1}{2}\omega\left[\varepsilon_{ij}^{ah} E_i^* E_j + \mu_{ij}^{ah} H_i^* H_j + \left(\zeta_{ij}^h - \xi_{ij}^h\right)\left(H_i^* E_j\right)^{ah} + \left(\zeta_{ij}^{ah} + \xi_{ij}^{ah}\right)\left(H_i^* E_j\right)^h\right]. \quad (12)$$

Evidently, both the average energy density and dissipative losses are defined not only by the ME-medium constitutive parameters, but also by the topological structure of the field determined by the dyadic $\vec{\vec{A}} \equiv \{E_i H_j^*\}$. This field structure, specific for ME media, we call a ME-field structure. In a case of a nondispersive ME medium, Eq. (11) describes the energy density of a magnetoelectrostatic structure [1 – 3].

Eqs. (11) and (12) were obtained with taking into account the basics of the tensor algebra. It is well known that a contraction of two second-rank Hermitian tensors is a real quantity. A contraction of two second-rank anti-Hermitian tensors is a real quantity as well. At the same time, a contraction of Hermitian and anti-Hermitian tensors of the second rank is an imaginary quantity. The conditions of losslessness of a ME medium ($\langle P \rangle = 0$), arise from a proper combination of constitutive parameters and parameters of the field dyadic $\vec{\vec{A}}$. One can see that a ME medium is definitely lossless, when $\varepsilon_{ij}^{ah} = \mu_{ij}^{ah} = 0$, $\zeta_{ij}^h = \xi_{ij}^h$, and $\zeta_{ij}^{ah} = -\xi_{ij}^{ah}$. These conditions can be rewritten as [57 – 59]

$$\vec{\vec{\varepsilon}}^T = \vec{\vec{\varepsilon}}^*, \quad \vec{\vec{\mu}}^T = \vec{\vec{\mu}}^*, \text{ and } \xi^T = \zeta^*. \quad (13)$$

When constitutive parameters satisfy relations (13), the ME energy density is expressed as

$$\langle W_{ME} \rangle = \frac{1}{4}\left\{\frac{\partial\left[\omega(\zeta_{ij}^h + \xi_{ij}^h)\right]}{\partial \omega}\left(H_i^* E_j\right)^h + \frac{\partial\left[\omega(\zeta_{ij}^{ah} - \xi_{ij}^{ah})\right]}{\partial \omega}\left(H_i^* E_j\right)^{ah}\right\}$$

$$= \frac{1}{2}\left[\frac{\partial(\omega \zeta_{ij}^h)}{\partial \omega}\left(H_i^* E_j\right)^h + \frac{\partial(\omega \zeta_{ij}^{ah})}{\partial \omega}\left(H_i^* E_j\right)^{ah}\right]. \quad (14)$$

Being not electromagnetic in nature, the energy density $W_{ME}$ plays a special role in forming a topological structure of the fields. The properties of the field structure arise from an analysis of Eq. (10). This equation can be transformed as follows:

$$\frac{E_{m_i}^*(t)}{H_{m_j}(t)} = \frac{\partial E_{m_i}^*(t)/\partial t}{\partial H_{m_j}(t)/\partial t} = \frac{dE_{m_i}^*(t)}{dH_{m_j}(t)}, \quad (15)$$

where $dE_{m_i}^*$ and $dH_{m_j}$ are differentials of the corresponding fields. The field constrains (10) are imposed simultaneously to all the field components and from Eq. (15) it is evident that there exists a linear time-relation coupling between complex amplitudes of the fields. In particular, it is fulfilled when both complex-amplitude vectors, $E_{m_i}$ and $H_{m_j}$, *rotate synchronically in time*. In a general form, we can represent the relation (15) as

$$E_{m_i}(t) = T_{ij} H_{m_j}^*(t), \quad (16)$$

where elements $T_{ij}$ of matrix $[T]$ are time-independent quantities. The matrix $[T]$ is a field-polarization matrix. This is an invariant defined for a specific type of ME medium. It is evident that components of matrix $[T]$ are complex quantities. *While in EM processes, electric and magnetic fields are interconnected (on the wavelength scale) due to Maxwell's equations, an intrinsic ME effect in a medium gives additional associations (on the subwavelength scale) between the fields described by the wave-polarization parameters $T_{ij}$.* To find parameters of the polarization matrix $[T]$, one should solve an electromagnetic boundary problem with the known medium constitutive parameters. When the parameters of matrix $[T]$ are found, an average energy density in a medium can be determined. In general, this is an integro-differential problem.

Let us consider, as a particular case, the transversal polarization matrix $[T_\perp]$ for the $x$, $y$ field components:

$$[T_\perp] \equiv \begin{bmatrix} \dfrac{E_{m_x}}{H_{m_x}^*} & \dfrac{E_{m_x}}{H_{m_y}^*} \\ \dfrac{E_{m_y}}{H_{m_x}^*} & \dfrac{E_{m_y}}{H_{m_y}^*} \end{bmatrix} = \begin{bmatrix} \dfrac{E_{m_x} H_{m_x}}{|H_{m_x}|^2} & \dfrac{E_{m_x} H_{m_y}}{|H_{m_y}|^2} \\ \dfrac{E_{m_y} H_{m_x}}{|H_{m_x}|^2} & \dfrac{E_{m_y} H_{m_y}}{|H_{m_y}|^2} \end{bmatrix}. \quad (17)$$

This is neither Hermitian nor anti-Hermitian matrix. Suppose that we have $|H_{m_x}| = |H_{m_y}| = |H_m|$. The matrix (17) can be rewritten as

$$[T_\perp] = \frac{1}{|H_m|^2} \begin{bmatrix} E_{m_x} H_{m_x} & E_{m_x} H_{m_y} \\ E_{m_y} H_{m_x} & E_{m_y} H_{m_y} \end{bmatrix}. \quad (18)$$

Evidently, the complex envelope of the transversal-field structure can have both the mutually parallel and perpendicular *x*, *y* components of the electric and magnetic fields. This analysis can be generalized for all the field components (transversal and longitudinal) of matrix $[T]$. It is worth noting also that for the components of matrix (18), we have a steady part

$$\langle E_i(t) H_j(t) \rangle = \frac{1}{2} \text{Re}\left[ E_{m_i}(\omega) H_{m_j}^*(\omega) \right] \tag{19}$$

together with a harmonically fluctuating part

$$E_i(t) H_j(t) = \frac{1}{2} \text{Re}\left[ E_{m_i}(\omega) H_{m_j}(\omega) e^{2i\omega t} \right]. \tag{20}$$

For an EM wave propagating in lossless non-ME media, the energy density associated with the electric field $W_E$ is equal to the energy density associated with the magnetic field $W_M$. So, the waves are characterized by synchronized oscillations of the electric and magnetic fields. In a case of ME media, the electric-magnetic duality is broken since the energy is not shared equally between the electric and magnetic fields. When, for an EM wave propagating in a lossless ME medium, the stored energy density $W$ is shared between the electric-field energy density $W_E$, the magnetic-field energy density $W_M$, and the ME energy density $W_{ME}$ (Fig. 1), the circulating processes of the energy interchange can be observed. Moreover, there could be non-reciprocal processes of the energy interchange – the processes with the time-symmetry breaking behaviors. In ME-medium electrodynamics, the fact that there exist three components of the energy density makes evident the presence of power flow circulation. There should be a *quasistatic circulation process in a subwavelength-domain region*. In this subwavelength domain, the energy exchange ought to occur slowly enough to remain in internal equilibrium. For a given electromagnetic-wave frequency, the ME-field power-flow circulation is an adiabatic process. It means that there should be no change of the energy of the system over the cycle. Another important aspect concerns the fact that these circulations can be quantized power-flow circulations.

What are the power flows of the circulating adiabatic processes of the energy interchange, shown in Fig. 1? Definitely, they are not exclusively the EM power flows. In our analysis of quasimonochromatic EM waves with complex amplitudes as smooth-fluctuation functions, we have to take into consideration also an additional propagation effect occurring on the subwavelength scale of an EM wave. Assuming that constitutive parameters are depended also on the wave vector [see Eqs. (4), (5)], we rewrite the energy balance equation (8) as [1, 19, 20, 55]

$$-\nabla \cdot \left( \langle \vec{S} \rangle + \langle \vec{A} \rangle \right) = \frac{\partial \langle W \rangle}{\partial t} + \langle P \rangle, \tag{21}$$

where $\langle \vec{A} \rangle$ is the average power flow density caused by the spatial dispersion. For ME medium we have:

$$\langle \vec{A} \rangle = -\frac{1}{4}\omega \left\{ \frac{\partial \varepsilon_{ij}^h}{\partial \vec{k}} E_i^* E_j + \frac{\partial \mu_{ij}^h}{\partial \vec{k}} H_i^* H_j + \frac{\partial \left( \zeta_{ij}^h + \xi_{ij}^h \right)}{\partial \vec{k}} \left( H_i^* E_j \right)^h + \frac{\partial \left( \zeta_{ij}^{ah} - \xi_{ij}^{ah} \right)}{\partial \vec{k}} \left( H_i^* E_j \right)^{ah} \right\}. \tag{22}$$

In Ref. [20], it was shown that Eqs. (21) and (22) can be obtained only if the following constraints to the slowly space-varying amplitudes are imposed:

$$E_{m_i}^*(\vec{r})\frac{\partial H_{m_j}(\vec{r})}{\partial \vec{r}} = H_{m_j}(\vec{r})\frac{\partial E_{m_i}^*(\vec{r})}{\partial \vec{r}}. \quad (23)$$

One can transform Eqs. (23) as follows:

$$\frac{E_{m_i}^*(\vec{r})}{H_{m_j}(\vec{r})} = \frac{\partial E_{m_i}^*(\vec{r})/\partial \vec{r}}{\partial H_{m_j}(\vec{r})/\partial \vec{r}} = \frac{dE_{m_i}^*(\vec{r})}{dH_{m_j}(\vec{r})}, \quad (24)$$

where $dE_{m_i}^*$ and $dH_{m_j}$ are differentials of the corresponding fields. Eq. (24) shows that there exist a linear space-relation coupling between complex amplitudes of the fields. In particular, it can be fulfilled when both complex-amplitude vectors, $E_{m_i}$ and $H_{m_j}$, *rotate synchronically in a space of the subwavelength domain*. In a general form, we represent the relation (21) as

$$E_{m_i}(\vec{r}) = R_{ij} H_{m_j}^*(\vec{r}), \quad (25)$$

where elements $R_{ij}$ of matrix $[R]$ are space-independent quantities, which are specified for a particular type of a ME medium and a concrete type of the boundary conditions.

The field properties in the $[T]$ matrix is also applicable to the $[R]$ matrix. It means that there are both *spin and orbital rotations* of the electric- and magnetic-field vectors in a subwavelength domain of the ME material structure. For EM processes in ME media, phase shifts between complex amplitudes of the electric and magnetic fields, $E_m \sphericalangle H_m$, can be not equal to $\pm 90°$ or $0° (180°)$ – usual phase shifts between complex amplitudes of the electric and magnetic fields in lossless EM processes in non-ME media. In topological subwavelength domains of ME media, one can observe the regions with both $\vec{E} \times \vec{H} \neq 0$ and $\vec{E} \cdot \vec{H} \neq 0$. We can distinguish the clockwise and counterclockwise topological-phase rotation of the ME fields. A vortex carries with it a certain angular momentum and energy. In an entire structure, we have a spontaneous alignment of the ME-field power-flow vortices. For propagating EM waves, the ME vortex is a type of a topological defect. The region of "ME kinetic energy" [defined by the term $\vec{r} \times (\vec{E} \times \vec{H})$] should be accompanied with the region of "ME potential energy" [defined by the term $\vec{E} \cdot \vec{H}$]. The ME-field power-flow vortex can hold energy in it rotation. Important to note that the quantity $\vec{E} \cdot \vec{H}$ is a pseudoscalar. Also, in non-enantiomeric strictures, the power-flow vortices should be due to time-reversal symmetry breaking. Assuming quantization of the orbital angular momentum, one can expect observation of quantization of ME energy.

From the above analysis of the energy balance conditions, we see that dynamic processes in ME media cannot be described sufficiently based only on constitutive relations (1), (2). The field constraints (10) and (23) should be imposed in addition. The effect is observed that the EM-wave propagation process in a ME medium is accompanied with topological-phase rotation of the fields in a subwavelength domain. Meanwhile, the following questions definitely arise: (i) What, in general, are these topological subwavelength domains? (ii) What kinds of discrete symmetry breakings can one observe on the walls of such domains? and, finally and most importantly, (iii) What is the nature of the rotating fields in subwavelength domains of ME materials – the ME fields?

### III. ME FIELDS AND QUASI-MAGNETOSTATIC QUANTIZED STATES

**1. Magnetic-dipolar-mode oscillations**

Considering magnetoelectricity as a magnetically driven phenomenon, we have to dwell now on the role of spin waves (magnons) in dynamical ME effects. In general, the theory of spin waves incorporates both the exchange and dipole-dipole interactions. A character of magnetic dipole-dipole ferromagnetism is essentially different from exchange ferromagnetism. The wavelength of dipole-dipole magnons [called also magnetostatic (MS) magnons] is much less than the EM wavelength but very large compared to the wavelength of short-range exchange-interaction magnons [1, 42 – 44]. Spin waves excited in an inhomogeneous modulated spin structure can be coupled to electric polarization via ferromagnetic-exchange interaction. In particular, dynamical ME coupling is observed in helical magnets with noncollinear spin configurations induced by exchange interactions [60, 61]. On the other hand, Khomskii pointed out [62] that in insulating ferromagnet with homogeneous modulated spin configuration, a spin-wave packet (of both the exchange and dipole-dipole interaction) will carry an electric-dipole moment directed perpendicular to the direction of the spin-wave propagation.

ME fields can be considered as quantized states of dipole-dipole magnons in subwavelength domains of ME materials. Such quantized states are well modeled by spectral characteristics of mode MDM oscillations observed in confined ferromagnetic-insulator structures. Space variations of these resonances are in the scales much less than the EM wavelength but very large compared to the exchange-interaction scales [44]. For EM-wave processes, a resonant region of MDM oscillations is viewed as a subwavelength domain. Assuming that this domain is of a cylindrical form, we can observe MDMs with azimuth variations. Nontrivial topology of MDM oscillations in such a sample manifests itself in unidirectionally rotating (chiral) eigenmodes. Breaking of symmetry between clockwise and counterclockwise modes due to the chiral boundary conditions results in appearance of MDM angular momenta and power-flow vortices [26, 53, 63, 64]. We have unidirectionally rotating spin-wave packet that carries an orbital angular momentum. The ME fields appear as a results of such complicated dynamics. The near-field structure outside of a MDM ferrite disk has regions with $\vec{E} \times \vec{H} \neq 0$ and the regions with $\vec{E} \cdot \vec{H} \neq 0$.

**2. MDM ferrite disks as ME particles**

In macroscopic electrodynamics, one can define three types of currents: density of the electric displacement current $\varepsilon_0 \frac{\partial \vec{E}}{\partial t}$, the electric current density arising from polarization $\frac{\partial \vec{p}}{\partial t}$, and the divergenceless electric current density arising from magnetization $\vec{\nabla} \times \vec{m}$ [1, 21]. For a magnetic insulator, we have the macroscopic Maxwell equation:

$$\vec{\nabla} \times \vec{B} = \mu_0 \left( \varepsilon_0 \frac{\partial \vec{E}}{\partial t} + \frac{\partial \vec{p}}{\partial t} + \vec{\nabla} \times \vec{m} \right) = \mu_0 \left( \frac{\partial \vec{D}}{\partial t} + \vec{\nabla} \times \vec{m} \right). \qquad (26)$$

In a case of a subwavelength magnetic sample (but with sizes that much exceed the sizes of the exchange-energy variation), one can neglect time variation of electric energy, stating that [1]

$$\left|\frac{\partial \vec{D}}{\partial t}\right| \ll \left|\vec{\nabla} \times \vec{m}\right|. \tag{27}$$

When a magnetic sample is at the condition of ferromagnetic resonance, we have extremely strong material temporal dispersion. In this case

$$\vec{B} = \mu_0 \left(\vec{H} + \vec{m}\right) \approx \mu_0 \vec{m}. \tag{28}$$

With use of conditions (27), (28), we can rewrite Eq. (26) as

$$\vec{\nabla} \times \vec{H} = 0. \tag{29}$$

These is a condition of quasi-magnetostatics oscillations in a magnetic insulator [1, 44]. The magnetic medium is anisotropic. So, the permeability is a tensor. In the frequency range of ferromagnetic resonance, components $\mu_{ik}$ of this tensor depend on the frequency. Following Eq. (29) we introduce the MS potential: $\vec{H} = -\vec{\nabla}\psi$. Taking into account that $\nabla \cdot \vec{B} = 0$, we have

$$\vec{\nabla} \cdot \left(\vec{\ddot{\mu}} \cdot \vec{\nabla}\psi\right) = 0 \tag{30}$$

This is the wave equation for the MS-wave function, called the Walker equation [65]. Outside a ferrite sample we have the Laplace equation $\nabla^2 \psi = 0$. We have a boundary-value problem with non-trivial solutions for certain values $\mu_{ik}$. These quantities determine eigenfrequencies of a non-homogeneous ferromagnetic resonance in a subwavelength sample [1, 44]. For a ferrite, magnetically saturated by a bias field $\vec{H}_0$ along axis $z$, the permeability tensor $\vec{\ddot{\mu}}\left(\omega, \vec{H}_0\right)$ has a form:

$$\vec{\ddot{\mu}} = \mu_0 \begin{bmatrix} \mu & i\mu_a & 0 \\ -i\mu_a & \mu & 0 \\ 0 & 0 & 1 \end{bmatrix}. \tag{31}$$

In a quasi-2D ferrite disk, with the disk axis oriented along $z$, the Walker-equation solution for the MS-potential wave function is written in a cylindrical coordinate system as [53, 63, 66]:

$$\psi = C\xi(z)\tilde{\varphi}(r,\theta), \tag{32}$$

where $\tilde{\varphi}$ is a dimensionless membrane function written in the $r$ and $\theta$ in-plane coordinates, $\xi(z)$ is a dimensionless function of the MS-potential distribution along $z$ axis, and $C$ is a dimensional amplitude coefficient. The MDMs in a ferrite disk are characterized by topologically distinct structures of the fields. Using electromagnetic boundary conditions, one obtains the following equation for a membrane wave function on a lateral surface of a ferrite disk of radius $\mathcal{R}$ [53, 63, 66, 67]:

$$\left(\tilde{\varphi}\right)_{r=\mathcal{R}^-} - \left(\tilde{\varphi}\right)_{r=\mathcal{R}^+} = 0$$

$$\mu\left(\frac{\partial \tilde{\varphi}}{\partial r}\right)_{r=\mathcal{R}^-} - \left(\frac{\partial \tilde{\varphi}}{\partial r}\right)_{r=\mathcal{R}^+} = -i\frac{\mu_a}{\mathcal{R}}\left(\frac{\partial \tilde{\varphi}}{\partial \theta}\right)_{r=\mathcal{R}^-}.$$

(33)

Evidently, in the spectral solutions, one can discern the time direction (given by the direction of the magnetization precession and correlated with a sign of $\mu_a$) and distinguish the azimuth rotation direction (given by a sign of $\frac{\partial \tilde{\varphi}}{\partial \theta}$). For a given sign of a parameter $\mu_a$, there are different MS-potential wave functions, $\tilde{\varphi}^{(+)}$ and $\tilde{\varphi}^{(-)}$, corresponding to the positive and negative directions of the phase variations with respect to a given direction of azimuth coordinates, when $0 \le \theta \le 2\pi$. So, a function $\tilde{\varphi}$ is not a single-valued function. It changes a sign when angle $\theta$ is turned on $2\pi$.

Taking into account Eqs. (33), the boundary-value-problem solution for MS-wave function in a qasi-2D ferrite disk is written as

$$\psi(r,\theta,z,t) = C_{\nu,n} J_\nu\left(\frac{\beta r}{\sqrt{-\mu}}\right)\left(\cos\beta z + \frac{1}{\sqrt{-\mu}}\sin\beta z\right)e^{-i\nu\theta}e^{i\omega t}.$$ (34)

Here $\beta$ is a wave number of a MS wave propagating in a ferrite along the $z$ axis, $\nu$ is a positive integer azimuth number, and $J_\nu$ is the Bessel function of order $\nu$ for a real argument. This equation shows that the modes in a ferrite disk are MS waves standing along the $z$ axis and propagating along an azimuth coordinate in a certain (given by a direction of a normal bias magnetic field) azimuth direction. For the known solution of the wave function $\psi(\vec{r},t)$, the power flow density is viewed as a current density [66, 67]:

$$\vec{\mathcal{J}} = \frac{i\omega}{4}\left(\psi\vec{B}^* - \psi^*\vec{B}\right).$$ (35)

This is a power flux arising from the dipole-dipole interaction of magnetic dipoles. Since MDMs in a ferrite disk are unidirectionally rotating (chiral) waves, circulation of vector $\vec{\mathcal{J}}$ along contour $L = 2\pi r$ is not equal to zero. Therefore, we can observe an orbital angular momentum due to the power-flow circulation:

$$\vec{L}_z = \oint_L \vec{r} \times \vec{\mathcal{J}}\, dl\ ,$$ (36)

This circulation is quantized.

When the spectral problem for the MS-potential scalar wave function $\psi(\vec{r},t)$ is solved, distribution of magnetization in a ferrite disk is found as $\vec{m} = -\vec{\chi}\cdot\vec{\nabla}\psi$, where $\vec{\chi}$ is the susceptibility tensor of a ferrite [44]. The magnetization has both the spin and orbital rotation. At the MDM resonance, the electric field in any point outside a ferrite disk is defined as [53, 63]

$$\vec{E}(\vec{r}) = -\frac{1}{4\pi}\int_V \frac{\vec{j}^{(m)}(\vec{r}')\times(\vec{r}-\vec{r}')}{|\vec{r}-\vec{r}'|^3}\,dV', \tag{37}$$

where $\vec{j}^{(m)} = i\omega\mu_0\vec{m}$ is the density of a magnetic current and frequency $\omega$ is the MDM-resonance frequency. Based on the known magnetization $\vec{m}$ inside a MDM resonator, one can find also the magnetic field distribution at any point outside a ferrite disk [21, 63]:

$$\vec{H}(\vec{r}) = \frac{1}{4\pi}\left(\int_V \frac{(\vec{\nabla}'\cdot\vec{m}(r'))(\vec{r}-\vec{r}')}{|\vec{r}-\vec{r}'|^3}\,dV' - \int_S \frac{(\vec{n}'\cdot\vec{m}(r'))(\vec{r}-\vec{r}')}{|\vec{r}-\vec{r}'|^3}\,dS'\right). \tag{38}$$

In Eqs. (37) and (38), V and S are a volume and a surface of a ferrite sample, respectively. Vector $\vec{n}'$ is the outwardly directed normal to surface S.

For the electric and magnetic fields defined by Eqs. (37) and (38), one can compose the vector

$$\langle\vec{S}_{MDM}\rangle \equiv \frac{1}{2}\mathrm{Re}(\vec{E}\times\vec{H}^*) \tag{39}$$

This vector has similar distributions as vector $\vec{\mathscr{J}}$, both inside and outside the ferrite disk. Circulation of vector $\langle\vec{S}_{MDM}\rangle$ along contour $L = 2\pi r$ is also similar to the power-flow circulation (36). The vector $\langle\vec{S}_{MDM}\rangle$ looks like a Poynting vector. Really, based on the vector relation $\vec{\nabla}\cdot(\vec{E}\times\vec{H}^*) = \vec{H}^*\cdot\vec{\nabla}\times\vec{E}^* - \vec{E}\cdot\vec{\nabla}\times\vec{H}^*$ with taking into account equations $\vec{\nabla}\times\vec{E} = -i\omega\vec{B}$, $\vec{H} = -\vec{\nabla}\psi$ and $\vec{\nabla}\cdot\vec{B} = 0$, one has as a result $\vec{\nabla}\cdot(\vec{E}\times\vec{H}^*) = i\omega\vec{\nabla}\cdot(\psi^*\vec{B})$. So, vector $\langle\vec{S}_{MDM}\rangle$ can be considered as a Poynting vector. This simple argumentation, used in literature [68], leaves aside, however, an important physical aspect. Compare to the case of EM wave propagation (with both curl electric and curl magnetic fields), we have here the modes with curl electric and *potential* magnetic fields. It is worth noting that there is no electromagnetic law describing transformation of the curl electric field to the potential magnetic field.

The angular momentum $\vec{L}_z$ is an intrinsic property of the fields at the MDM resonances, unrelated to the rigid-body rotation of a ferrite-disk. Because of the spin-orbit interaction between the spin and orbital rotation angular momenta of magnetization, there exists another quadratic relationship between the electric and magnetic field components in a near-field structure, in addition to the Eq. (39). It was shown [26, 53, 63, 64, 66] that outside a MDM ferrite sample there is also the region with

$$F = \frac{\varepsilon_0}{4} \text{Im}\left[\vec{E} \cdot \left(\nabla \times \vec{E}\right)^*\right] = \frac{\omega \varepsilon_0 \mu_0}{4} \text{Re}\left(\vec{E} \cdot \vec{H}^*\right) \neq 0. \tag{40}$$

We call the pseudoscalar parameter $F$ the helicity of a ME field. The helicity factor distributions for the main MDM resonance in the vacuum near-field regions above and below the disk, is shown in Fig. 2. The factor $F$ is characterized by antisymmetrical distribution with respect to disk axis. Along with this, the helicity factor changes a sign at time reversal. The region with positive helicity factor $F^{(+)}$ is a region with positive ME energy, $W_{ME}^{(+)}$. In this region, the orbital angular momentum $\vec{L}_z$ is directed along $z$ axis. The region with negative helicity factor $F^{(-)}$ is a region with negative ME energy, $W_{ME}^{(-)}$. In this region, the orbital angular momentum $\vec{L}_z$ is directed along $-z$ axis. This explains the picture of power flow rotations shown in Fig. 3. The "source" of the helicity factor is the pseudoscalar quantity of the magnetization distribution in a ferrite disk at the MDM resonances [64]:

$$\int_V \vec{m} \cdot \left(\vec{\nabla} \times \vec{m}\right)^* dV \neq 0, \tag{41}$$

$V$ is a volume a ferrite sample. These magnetization parameters are distributed asymmetrically with respect to the $z$ axis (see Fig. 4). Thus, the distribution of the helicity factor is also asymmetric.

The near fields of a MDM ferrite disk are characterized by the violation of the invariances under space reflection and time inversion – necessary conditions for emergence of the ME fields. From Eqs. (39) and (40), one can see that there exist both the regions with $\vec{E} \times \vec{H} \neq 0$ and the regions with $\vec{E} \cdot \vec{H} \neq 0$. The power-flow vortices are characterized by a topological charges. Energy for the power flow rotation arises from the ME energy determined by the helicity factor $F$.

## IV. ENERGY QUANTIZATION IN SUBWAVELENGTH ME DOMAINS

By modeling subwavelength domains of ME materials by MDM ferrite-disk resonators, we can see that the magnetic, electric, and ME energies of these structural elements are quantized quantities. In MDM resonances, an energy interchange should occur between these quantities. For a ferrite disk embedded into free space, we have the "helicity neutrality": $F^{(+)} = -F^{(-)}$. It means that in the near-field region we have $\int_{V^{(+)}} W_{ME}^{(+)} dV = \int_{V^{(-)}} W_{ME}^{(-)} dV$. For the total ME energy in a near-field region, we can write

$$\int_{V^{(+)}+V^{(-)}} W_{ME} dV = \int_{V^{(+)}} W_{ME}^{(+)} + \left(-\int_{V^{(-)}} W_{ME}^{(-)}\right) = 2 \int_{V^{(+)}} W_{ME}^{(+)} dV. \tag{42}$$

In a lossless structure, the total "ME potential energy" should be equal to the "ME kinetic energy" of the power-flow rotation:

$$\int\limits_{V^{(+)}+V^{(-)}} W_{ME} dV = \int\limits_{V^{(+)}+V^{(-)}} W_{rotation} dV . \qquad (43)$$

In a quasistatic process of energy circulation in a subwavelength ferrite particle, the energy density of the electric field must also be included, together with the magnetic-field energy density and the ME energy density. However, when solving the eigenvalue problem for MS-potential function $\psi$, we completely neglect the role of electric polarization, properly assuming that in the magnetization dynamics, time variation of electric energy is negligibly small and therefore the electric field in a subwavelength magnetic sample is very weak [1]. Nevertheless, the requirement for smallness of the electric field is not necessary to satisfy condition (27). The inequality (27) can also occur when the electric energy is not very small compared to the magnetic energy, but there are no time variations of vectors $\vec{E}$ and $\vec{p}$ with respect to vector $\vec{m}$ (and, certainly, with respect to space derivatives of vector $\vec{m}$). It means, in other words, that the lines of the electric field $\vec{E}$ as well the lines of the polarization $\vec{p}$ are "frozen" in the lines of magnetization $\vec{m}$. Such a condition can be fulfilled when both vectors $\vec{D} \equiv \varepsilon_0 \vec{E} + \vec{p}$ and $\vec{m}$ *rotate synchronically in time*. For orbitally rotating field patterns, originated from the magnetization dynamics in a ferrite disk, the MS description is valid even taking into account electric polarization, since no displacement electric current is still considered. At MDM resonances in a qiasi-2D ferrite disk, electric polarization is an observable quantity due to the orbital angular momentum of the magnetization. In this case, electric charges arise on the disk planes. This becomes clear from the following consideration.

Since a function $\tilde{\varphi}$ is not a single-valued function (it changes a sign when angle $\theta$ is turned on $2\pi$), for any mode $n$, the function $\tilde{\varphi}_n$ is a two-component sprinor pictorially denoted by two arrows:

$$\tilde{\varphi}_n^{\uparrow\downarrow}(\vec{r},\theta) = \begin{bmatrix} \tilde{\varphi}_n^{\uparrow} \\ \tilde{\varphi}_n^{\downarrow} \end{bmatrix} = \tilde{\eta}_n(\vec{r},\theta) \begin{bmatrix} e^{-\frac{1}{2}i\theta} \\ e^{+\frac{1}{2}i\theta} \end{bmatrix} \qquad (44)$$

Circulation of gradient $\vec{\nabla}_\theta \tilde{\varphi}$ along contour $L = 2\pi r$ is not equal to zero. When rotating around this contour, half-quantum behavior is observed [67].

For eigenfunctions $\tilde{\eta}$, we have conditions of orthogonalization of magnetic energy [26, 67]:

$$\left[ (W_M)_n - (W_M)_{n'} \right] \int\limits_{S_c} \tilde{\eta}_n \tilde{\eta}_{n'}^* dS = 0 . \qquad (45)$$

The spectral problem for membrane function $\tilde{\eta}$ is formulated by applying the Neumann-Dirichlet boundary conditions

$$\left( \tilde{\eta}_n \right)_{r=\mathcal{R}^-} - \left( \tilde{\eta}_n \right)_{r=\mathcal{R}^+} = 0 \qquad (46)$$

and

$$\mu\left(\frac{\partial \tilde{\eta}_n}{\partial r}\right)_{r=\mathcal{R}^-} - \left(\frac{\partial \tilde{\eta}_n}{\partial r}\right)_{r=\mathcal{R}^+} = 0, \tag{47}$$

which differ from the electromagnetic boundary conditions (33) used for spectral solutions for functions $\tilde{\varphi}$.

The spectral problem for functions $\tilde{\varphi}$ can be represented as a result of the spin-orbit interaction of eigenfunctions $\tilde{\eta}$ with a certain surface magnetic current $j_s^{(m)}$ circulating on a lateral surface of the disk. This current should compensate the term $\left(i\mu_a \frac{1}{r}\frac{\partial \tilde{\varphi}_n}{\partial \theta}\right)_{r=\mathcal{R}^-}$ in the boundary conditions (33). One can see that for a given direction of a bias magnetic field (that is, for a given sign of $\mu_a$), there are two, clockwise and counterclockwise, quantities of a circulating magnetic current. The current $j_s^{(m)}$ is defined by the velocity of an irrotational border flow. This magnetic current is a divergenceless bound current, not having any charge accumulation associated with it. This type of persistent current appears due to a mesoscopic effect: the magnitude of the current becomes appreciable when the size of the ferrite system is reduced to the scale of the dipole-dipole quantum phase coherence length of precessing electrons.

At the MDM resonance, the magnetic chiral current $\vec{j}_s^{(m)}$ on a lateral surface of a ferrite disk results in appearance of an anapole moment [67, 69]

$$a_{\pm}^{(e)} \propto \mathcal{R} \int_0^d \oint_{\mathcal{L}} \left[\vec{j}_s^{(m)}(z)\right]_\theta \cdot d\vec{l}dz. \tag{48}$$

This anapole moment, being time reversal odd and parity odd, leads to emergence of DC electric charges on the ferrite-disk planes [69].

A ferrite disk is made of a magnetic dielectric with a sufficiently high dielectric constant. For example, the yttrium iron garnet (YIG) is an isotropic dielectric with a dielectric constant of $\varepsilon_r = 15$. By virtue of the electric field originated from the magnetization dynamics [see Eq. (37)], every separate electric dipole in a ferrite disk precesses around its own axis. While in precession motion, electric dipoles accomplish also an orbital geometric-phase rotation. Thus, a ferrite disk is subjected with an induced electric gyrotropy and orbitally driven electric polarization. Orbital torques exerting on the electric polarization due to a rotating MDM electric field should be equal to reaction torques exerting on the magnetization in a ferrite disk. Because of this reaction torques, the precessing magnetic moments of the ferromagnet will be under additional mechanical rotation at a certain frequency $\Omega$. This frequency is defined based on both, spin and orbital, angular momentums of the fields of MDM oscillations. As a result, the Larmor frequency of a ferrite-disk structure will be lower than such a frequency in a ferrite sample without the orbitally rotating fields [26, 53]. Thus, due to electric polarization, one observes the Doppler shift at the MDM resonances. Orbitally rotating magnetization simulate the electric fields, which, in turn, stimulate the electric dipoles that affect the magnetic structure.

Fig. 5 shows the energy distribution in a ferrite disk at the MDM resonance. The magnetic-field energy $W_M$ provides the ME energy $W_{ME}$ for the field rotations. Due to this rotations we observe the electric-field energy $W_E$. Because of the reaction torque, the precessing magnetic moments in a ferromagnet are under additional mechanical rotation. It results in change of the magnetic-field energy density $W_M$. This circulating process of the energy interchange is non-reciprocal. It is the process with the time-symmetry breaking behavior.

The energy circulation processes considered above, can be observed in a subwavelength-domain regions of a ME material. In EM-wave processes in a lossless non-ME medium, we have conversion of electric energy to magnetic energy and vice versa. In a case of a lossless ME medium, there are mutual conversions between the electric and ME energies, as well as mutual conversion between the magnetic and ME energies. Since the energy density $W_{ME}$ is non-electromagnetic in nature, we have, in fact, a combined effect of the EM and ME dynamics. An essential factor in this case is that ME fields can also be characterized by reactive helicity parameters and reactive power flows [70]. For EM wave, propagating in a ME medium, the Poynting vector is complex. It means that together with the energy balance equation for the time-averaged values (9), there should be also the equation for an imaginary part of the Poynting vector. Neglecting losses, we have [21]:

$$\mathrm{Im}\int_V \left(\vec{j}\cdot\vec{E}\right)_{extraneous} dV + 2i\omega\int_V \left(W_E - W_M\right) dV + \mathrm{Im}\oint_S \vec{S}\cdot\vec{n}\, dS = 0, \tag{49}$$

where $\vec{S} = \frac{1}{2}\left(\vec{E}\times\vec{H}^*\right)$ is a complex Poynting vector, $\mathrm{Im}\int_V \left(\vec{j}\cdot\vec{E}\right)_{extraneous} dV$ is an average reactive power of a source and $\mathrm{Im}\oint_S \vec{S}\cdot\vec{n}\, dS$ is an average reactive power passing through a surface $S$. The presence of an extraneous (non-electromagnetic) source leads to the fact that ME medium is not transparent for EM radiation. The presence of a second harmonic term in Eq. (49) is definitely related to the half-quantum circulations observed at the MDM resonances [67].

## V. A NOTE ON NATURAL AND ARTIFICIAL MAGNETOELECTICITY

When we are talking on magnetoelectric dynamics, we have to refer also to artificial structures – bianisotropic metamaterials. The notion "bianisotropic media" had been introduced to generalize different effects of coupling between magnetic and electric properties [71]. It is assumed that this notion embraces both the ME interaction phenomena in crystals and cross polarization effects in particulate composites – metamaterials. Bianisotropic metamaterials are characterized by constitutive relations (4), (5). With such constitutive parameters, so-called chiral and Tellegen metamaterials are considered as particular cases of bianisotropic metamaterials. The local bianisotropic media is supposed as the media composed by structural elements with "glued" pairs of electric and magnetic dipoles. Accounting quadrupole and high-order multipole transitions represents, in fact, an accounting of spatial dispersion [72, 73].

It is assumed that bianisotropy (chirality) in these metamaterials arises from a local ME effect [74 – 77]. Such a "first-principle", "microscopic-scale" ME effect of a structure composed by "glued" pairs of electric and magnetic dipoles raises many questions. Firstly, it is unclear what the *near field* of this structure is. In metamaterial bianisotropic (chiral) structures, experimental retrieval of the cross polarization parameters are via far-field measurement of the scattering-matrix characteristics [78 – 80]. Retrieved permittivity and permeability frequently retain non-physical values, especially near the metamaterial resonances where most interesting features are expected. Far-field retrieved cross polarization parameters of "bianisotropic particles" retain much greater non-physical value. In classical electrodynamics, subwavelength electric dipole can be created by "microscopic-scale" linear electric current, while subwavelength magnetic dipole can be created by "microscopic-scale" circular electric current. In an artificial structure, the combined effect of these two types of "microscopic-scale" currents can be observed experimentally only in a far-field region. The far-field phenomena of bianisotropy (chirality) are absolutely not related to the near-field manipulation effects. The effects of Rayleigh scattering and cross polarization are incompatible.

Obviously, in bianisotropic metamaterials, no magnetoelectrostatic limit [see Eqs. (7), (8)] can be assumed. Since the electric and magnetic dipoles do not interact electromagnetically in the subwavelength region [21], there are no effects related to accumulation of ME energy $W_{ME}$. It is insufficient to characterize ME medium as a special medium with just only a cross-polarization effect. Together with constitutive relations (4), (5), certain constraints should by imposed to the field structure. The energy balance equation reveals unusual topological structure of the local fields in ME materials. To observe the ME energy in a near field region, one has to show the existence of the fist-principle near fields – the ME fields. The ME fields are not Maxwellian near fields. These are subwavelength-domain fields with specific properties of violation of spatial and temporal inversion symmetry.

## VI. CONCLUSION

In an analysis of EM wave propagation in a ME medium, it is insufficient to take into account just only a cross-polarization effect in this medium. Together with local constitutive parameters certain constraints should by imposed to the local field structure. The energy balance equation reveals unusual topological structure of the local fields in ME materials. These fields, called ME fields, are subwavelength-domain fields with specific properties of violation of spatial and temporal inversion symmetry. We argue that ME energy in EM systems can appear due to local (subwavelength) fields rotated adiabatically in unit cells of ferrotoroidic domains. ME fields can be considered as quantized states of magnetostatic magnons in subwavelength domains of ME materials. Such quantized states are well modeled by spectral characteristics of magnetic-dipolar-mode oscillations observed in confined ferromagnetic-insulator structures. Quantized ME fields arising from nonhomogeneous ferromagnetic resonances with spin-orbit effect, suggest a conceptually new microwave functionality for material characterization. In a view of our analysis, the "first-principle", "microscopic-scale" ME effect in bianisotropic metamaterials raises questions on the near-field topology and accumulation of ME energy in such material structures composed by "glued" pairs of electric and magnetic dipoles.

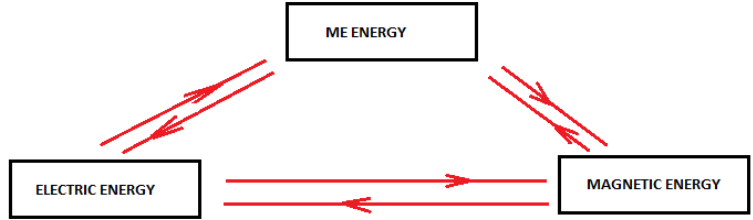

Fig. 1. For an EM wave propagating in a lossless ME medium, the stored energy density $W$ is shared between the electric-field energy density $W_E$, the magnetic-field energy density $W_M$, and the ME energy density $W_{ME}$. One can observer a quasistatic circulation process of the energy interchange in a subwavelength-domain region. There could be non-reciprocal processes of power flow circulations with time-symmetry breaking behaviors.

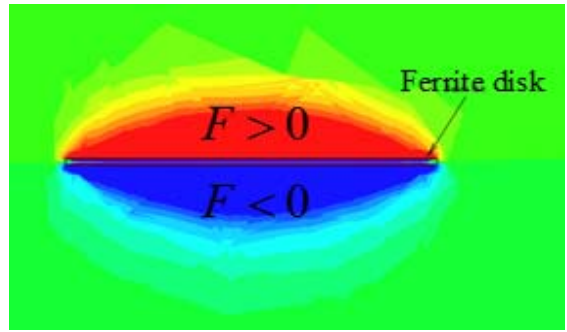

Fig. 2. The helicity factor for a MDM resonance in the vacuum near-field regions above and below a ferrite disk. The factor $F$ is characterized by antisymmetrical distribution with respect to the disk axis. Along with this, the helicity factor changes a sign at time reversal. A red region, where $F > 0$, we conditionally call $F^{(+)}$ region. This is the region with positive ME energy, $W_{ME}^{(+)}$. A blue region, where $F < 0$, we call $F^{(-)}$ region. This is the region with negative ME energy, $W_{ME}^{(-)}$. The regions $F^{(+)}$ and $F^{(-)}$ have volumes $V^{(+)}$ and $V^{(-)}$, respectively. A green region, where $F = 0$, is the region of regular EM fields.

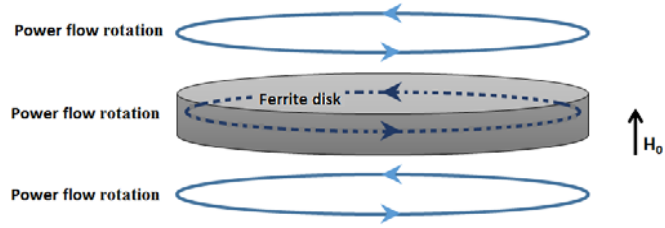

Fig. 3. A schematic picture of power flow rotations observed at MDM resonances. For a given direction of the bias magnetic field, a character of power-flow circulations is the same inside a ferrite and in the vacuum near-field regions above and below the disk.

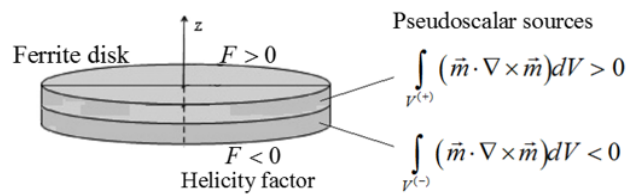

Fig. 4. Pseudoscalar quantity of the magnetization in a ferrite disk as a "source" of the helicity factor at the MDM resonance.

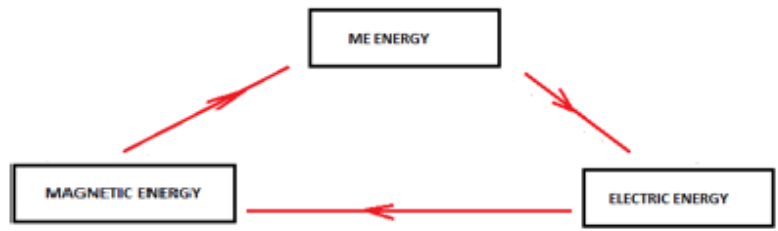

Fig. 5. The energy distribution in a ferrite disk at the MDM resonance. The magnetic-field energy $W_M$ provides the ME energy $W_{ME}$ for the field rotations. Due to this rotations we observe the electric-field energy $W_E$. Because of a reaction torque, the precessing magnetic moments in a ferromagnet are under additional orbital rotation. It results in change of the magnetic-field energy density $W_M$. This circulating process of the energy interchange is non-reciprocal. There is the process with the time-symmetry breaking behavior.